\documentclass[12pt,a4paper,final]{iopart}
\usepackage{iopams}
\usepackage{graphicx}
\usepackage[breaklinks=true,colorlinks=true,linkcolor=blue,urlcolor=blue,citecolor=blue]{hyperref}
\usepackage[labelfont=bf]{caption}
\usepackage[usenames,dvipsnames,svgnames,table]{xcolor}
\usepackage{enumerate}
\usepackage{changepage} 
\usepackage{multirow}	
\usepackage{soul}       
\usepackage{ulem}       
\usepackage{xcolor}     
\usepackage[hang]{footmisc}	

\setstcolor{red}    

\begin{document}
\title{Gender gap and polarisation\\of physics on global courses}

\author{Allan L. Alinea}
\address{Department of Physics, Osaka University, Japan}
\ead{alinea@het.phys.sci.osaka-u.ac.jp}

\author{Wade Naylor}
\address{School of Education, University of the Sunshine Coast, QLD, Australia}
\address{Department of Physics, Osaka University, Japan}
\ead{wnaylor@usc.edu.au}

\begin{abstract}
\noindent 
We extend on previous research on the Force Concept Inventory (FCI) given to first year classical mechanics students ($N=66$ students, over four years) pre and post score, for students on an international (global) course at Osaka University. In particular, we revisit the notion of ``polarisation'' in connection with the six polarisation-inducing questions in the FCI and examine its gender aspect. Our data suggest that this phenomenon is not unique to one gender. Furthermore, the extent by which it is exhibited by males may differ from that of females at the beginning (pretest) but the gap closes upon learning more about forces (posttest). These findings are for the most part, complemented by our result for the FCI as a whole. Although the differences in means for males and females suggest a gender gap, statistical analysis shows that there is no gender difference at the $95\%$ confidence level.
\end{abstract}

\vspace{2pc}
\maketitle
\tableofcontents

\section{Introduction}
It can be said that the Newton's laws of motion, the three pillars of Newtonian mechanics, are all about force. This is the reason why in a class on (elementary) Newtonian mechanics, much of the quantitative and conceptual problems that students are expected to solve involve friction, gravity, tension, weight, etc., in connection with elevators, pendula, bridges, projectiles, and other physical systems. In all of these problems, when it comes to dynamics, everyone essentially starts with the basics of Free Body Diagram (FBD). And when it comes to FBDs, everybody begins with identifying forces: ``What are the forces acting on a given body?'' This is so basic. Yet the basic skill of identifying forces reverberates up to the pinnacle of complexity involved in building bridges and towers, constructing satellites, and creating machineries in factories, among others. In the Force Concept Inventory (FCI) \cite{Hallouin, FCI, Hestenes}, regarded as the \textit{de facto} standard for concept inventories in Newtonian mechanics, six (that is, 20\%) of the 30 questions basically ask only one thing: identify the forces acting on a given body. It only shows that the importance of learning this basic skill cannot be overemphasised.

In our previous work \cite{alalinea}, we focused on the six FCI questions\footnote{These are the sets of questions \{5, 11, 13, 18, 29, 30\} in the FCI.} and studied how our students on global courses at Osaka University (Japan) \cite{CBCMP} answered them. We found out a ``phenomenon'' called \textit{polarisation}. To wit, given five choices corresponding to five sets of forces acting on a body specified in an FCI question, the great majority of students elected only two letters --- the \textit{polarising choices}. These choices consist of two sets of forces with one being a subset of the other. The subset in the pair is the correct answer while the other one includes extra ``fictitious'' force not legitimately acting on the body. Students who are otherwise knowledgeable about the acting forces are confused with the addition of the wrong force; thus, resulting to polarisation. We explained in our work that such a ``phenomenon'' may be attributed to misleading \textit{ontological categorisation}\cite{Johnston}. In the concluding remarks, as part of our future prospect, we opened the door to further investigation of this matter --- causes, cure, and other aspects of polarisation.

The current paper is a continuation of the previous work we mentioned above. Here, focusing on the possible \textit{other aspect} of polarisation, we investigate \textit{gender gap} \cite{MadsenRev, Bates, Docktor, Coletta, Glaser}. Gender gap is the resulting gender difference in favour of males in the student performance on standardised assessments such as the FCI. Such a delicate concept has so far found no simple explanation and complete ``cure'' that are supported by repeatable experiments. We are not here going to pretend to fill in the absence of a complete explanation and ``cure''. Our aim is more modest and being such, should be answerable at least within the range of validity of the available data that we have. We want to know if there is a gender aspect associated with polarisation. To this end, we wish (a) to determine whether polarisation is unique to one gender, and if it is not, (b) to establish if there is a gender difference associated with the extent by which polarisation is exhibited by students. 

\begin{figure}[tbh!]
	\centering
	\includegraphics[scale=0.95]{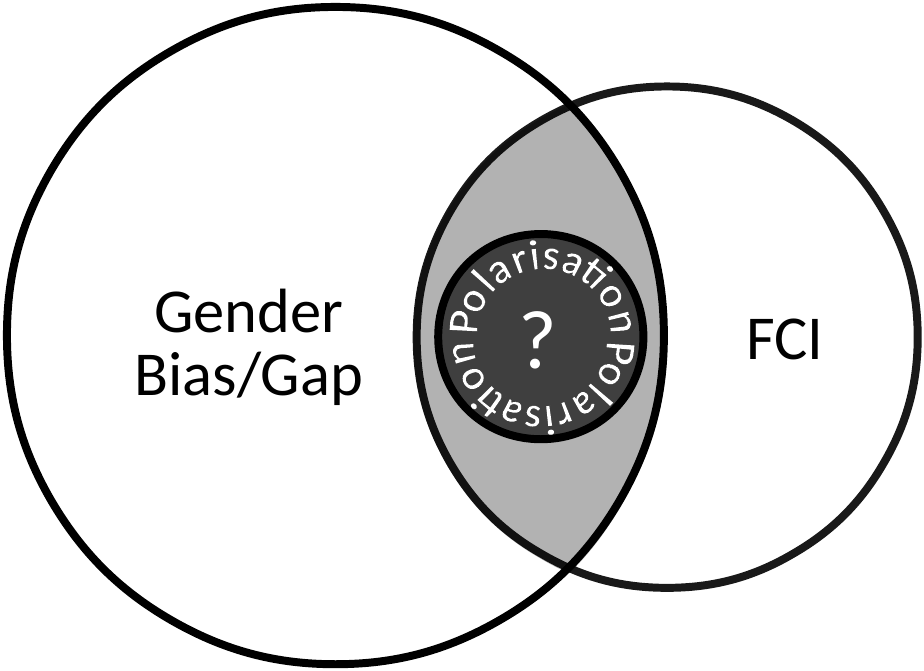}
	\caption{Venn diagram for polarisation, gender bias/gap, and the FCI.}
	\label{fig_venn_expect}
\end{figure}

Graphically, these objectives can be represented in a unified way through the placement of the circle for polarisation in the Venn diagram shown in \textbf{Fig. \ref{fig_venn_expect}}. Considering that the gender gap/bias extends beyond the FCI (eg., in other concept inventories, gender bias on Wikipedia \cite{wikiGenderBias}, etc), we have drawn a bigger circle for it and made it partially intersect with the circle for the FCI. Such a \textit{partial} intersection is justified, noting that it may only be \textit{partially} gender bias, if ever it is. With the six polarisation-inducing questions within the FCI, we want to know, where the corresponding circle for polarisation should be placed. Depending on our result, it may be completely moved out of the light gray-shaded region, meaning no gender gap, or it may be pushed a little further from its current location, partially intersecting with the big circle for gender bias/gap.

\section{Methodology}
\label{Meth}
The subjects of our testing are the students of the Chemistry-Biology Combined Major Program (CBCMP). This is a four-year program offered by Osaka university (Japan), open to all qualified students from all over the globe. Approximately 90\% of our students came from outside Japan. With English as the medium of instruction, all applicants are required to have sufficient English ability to be accepted. Considering all entrants, the average equivalent TOEFL score\footnote{We performed the necessary conversion for some students who took IELTS. Furthermore, we assigned full score for native speakers.} is 104 with a standard deviation of 14. Such a high average with relatively low standard deviation suggests that English ability is not a factor that could significantly affect performance on the FCI; this, we found out in our previous work \cite{alalinea}.

In this work, we have updated our data to include the batch for the Academic Year (AY) 2014-15. In total, starting from AY 2011-12, we have 66 students (56\% females and 44\% males) corresponding to four batches of entrants. These students attended calculus-based introductory Newtonian mechanics class under our supervision. Each class meeting lasts for 1.5 hours (90 minutes) and is conducted once a week for a period of four months\footnote{This relatively tight schedule is a common setting for university lecture classes in Japan.}. A typical class starts with a discussion of pre-lecture conceptual questions as refresher. These questions are usually given as assignment one meeting before the class to prepare the students. A short lecture then follows highlighting the main points of the subject for the day. After this, for the remaining 50 minutes, students are paired\footnote{See Refs. \cite{smith, Fagen} and the references therein for the benefits of peer discussion/instruction.} to answer questions on MasteringPhysics${}^\textnormal{\footnotesize TM}$ \cite{masteringPhys}. While solving conceptual and quantitative problems, the teaching assistant and the instructor go around the room to attend to student concerns about the problem they are solving. The students are given one week to finish the assigned task.

We administered the FCI at the beginning and end of every second semester from AY 2011-12 to AY 2014-15. All student answers for each question were recorded and an item analysis was performed with emphasis on the six polarisation-inducing questions. Standard statistical parameters for the four-year data were calculated including \textit{p}-value for \textit{t}-test, mean, standard deviation, percent difference, etc. Using these parameters, (a) student performance from pretest to posttest, (b) possible gender difference in overall FCI, (c) existence of polarisation, (d) possible gender difference in the polarization-inducing questions, among others, were analysed.

\section{Polarisation}
\subsection{Polarisation as a Whole}
\begin{figure}[tbh!]
	\makebox[\textwidth][c]{
		\includegraphics[width=1.2\textwidth]{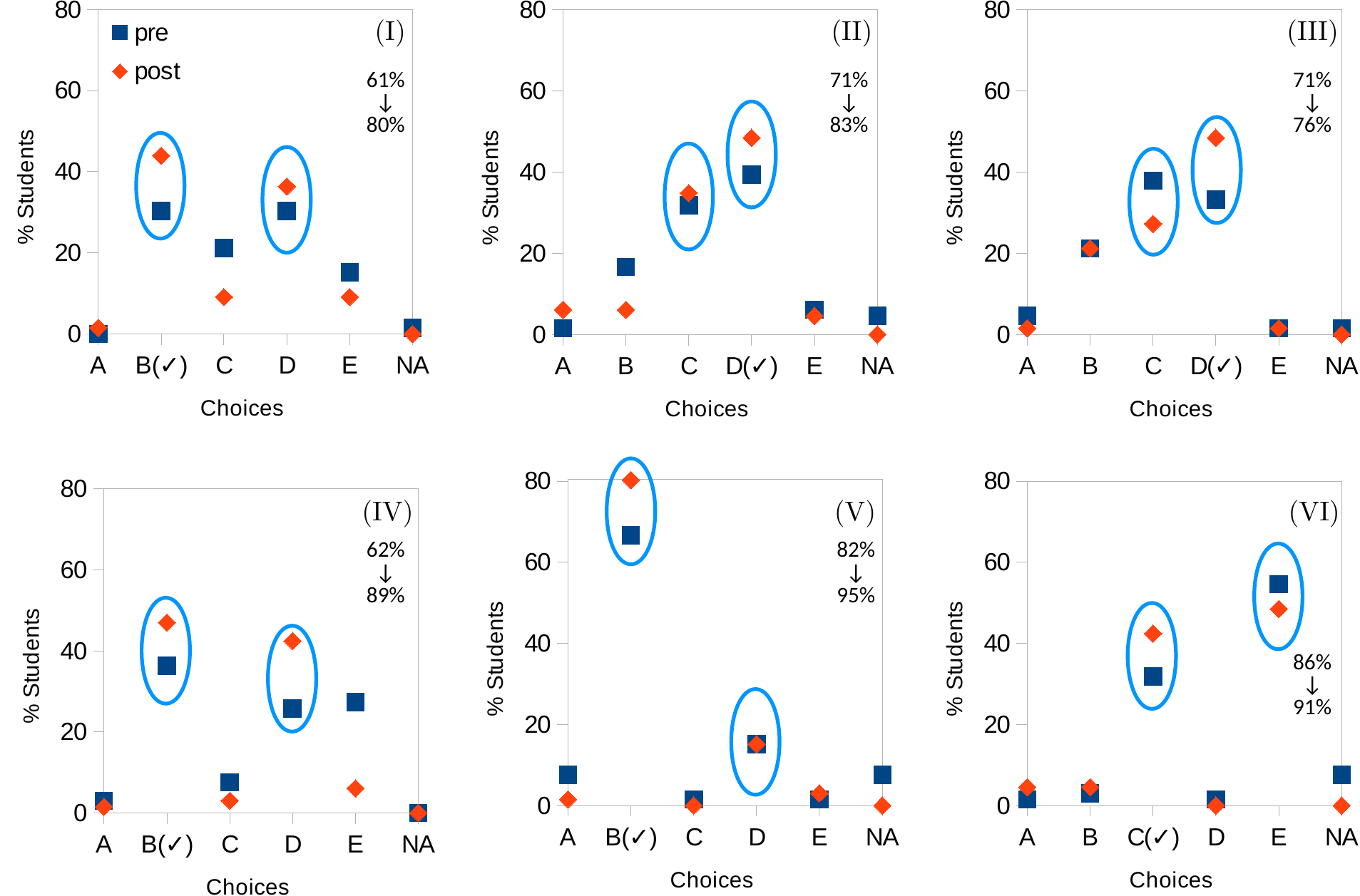}
	}
	\caption{Distribution of student answers for questions I to VI. The set of five choices spans from letters A to E, with NA corresponding to no answer. The correct answer for each question is labeled with a check mark. For each question, two letters (including the correct answer) take the majority share of student answers (see the encircled data points). The two numbers at the right part of each graph indicate the percent of students who chose these so-called \textit{polarising choices} from pretest to posttest.}
	\label{fig_a_to_e}
\end{figure}

Before we delve into the issue of the possible gender aspect of polarisation, we look at the extent of this phenomenon for all students irrespective of gender. \textbf{Figure \ref{fig_a_to_e}} shows the pair of dominant answers for questions I to VI. This pair as what we pointed out in our previous work \cite{alalinea}, is composed of a correct choice (CC) and a wrong choice (X), where CC happens to be a subset of X; e.g., in question I, these are choices (CC, X) = (B,D). Here, with our updated data, we once again confirm the observation that majority of students only select two choices. For the pretest at least 61\% with a maximum of 86\% elected the polarising choices. It rose to at least 76\% in the posttest with an astonishing maximum of 95\%. 

\begin{figure}[tbh!]
    \centering
    \includegraphics[scale=0.95]{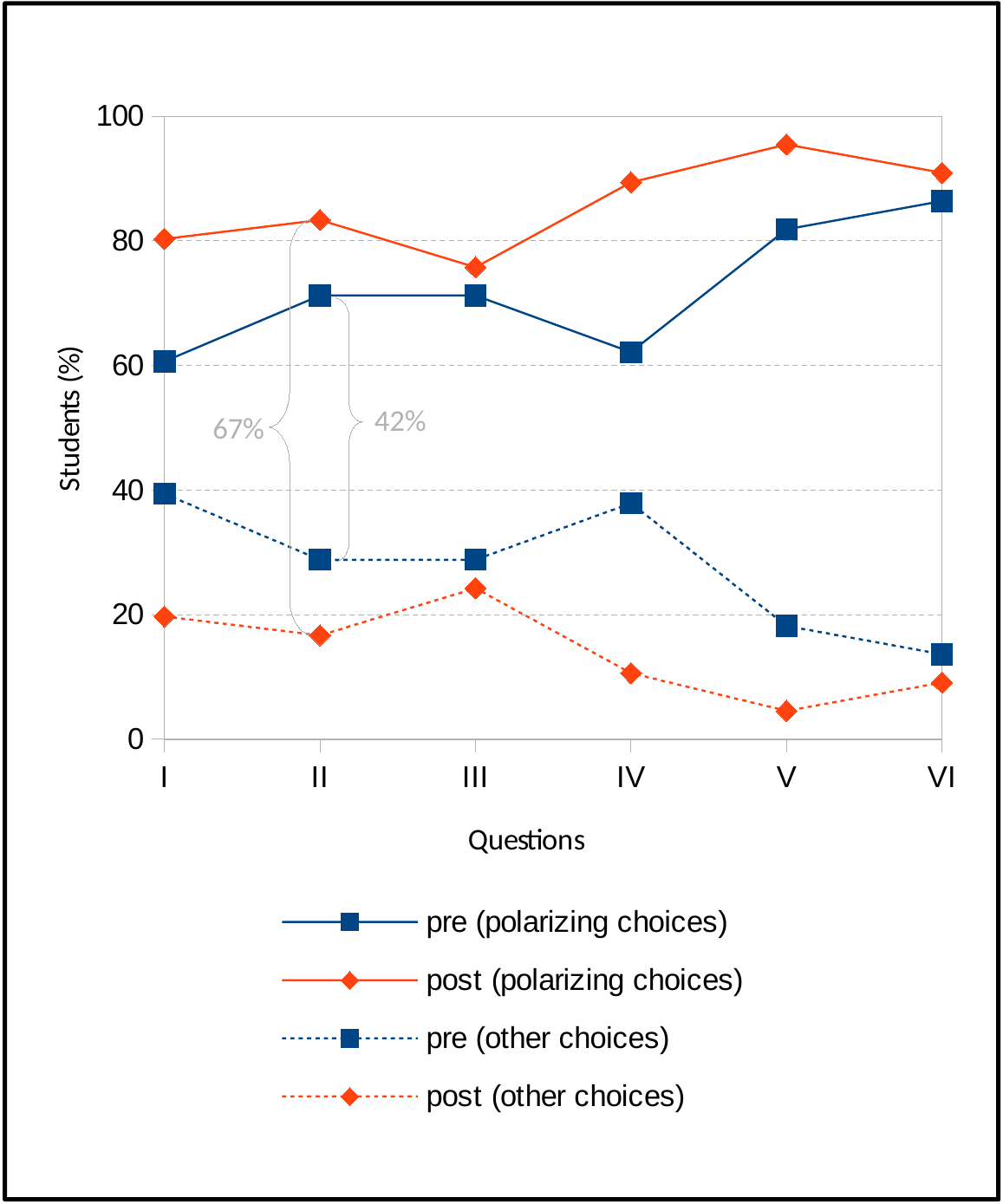}
    \caption{Distribution of students who elected the \textit{polarising} choices (solid lines) over the \textit{other} choices (dashed lines) in the six FCI questions I to VI.}
    \label{fig_pol_percent}
\end{figure}

\textbf{Figure \ref{fig_pol_percent}} shows a vivid graphical display of such polarisation from pretest to posttest. As can be seen in the figure, overall, there is a general increase in the proportion of students who elected the polarising choices from pretest to posttest. In tandem with \textbf{Fig. \ref{fig_a_to_e}}, we see that such a trend is strictly followed for all questions for the choice CC, and moderately obeyed for the choice X (see graphs for questions III and VI). A detailed examination of student responses indicates that the increase in the share for CC has a major contribution due to the migration of students who elected X in the pretest to CC in the posttest; that is, from being ``confused'' to ``enlightened''. More specifically, of all the transitions from the four wrong choices to CC, 54\% came from X $\rightarrow$ CC. Students who were once confused by the ``fictitious'' force in X, learned along the way, and elected CC in the posttest. This however, did not in general, decreased the share for X. Some students who elected the \textit{other} (three) choices in the pretest, somehow learned along the way, and chose X in the posttest. 

This behavior gives us a bird's eye view of the way student learns to identify forces acting on a given body. The dominant transition goes from the \textit{other} choices to the \textit{polarising} choices. Then within polarising choices, the transition goes from X to CC. This double-transition behavior resulted in the general increase in the proportion of students who elected the polarising choices from pretest to posttest. We view such a rise in the share for polarising choices whether it be due to choosing CC or X, as an indication of learning. Electing (the slightly correct choice) X although it is wrong, signifies that a student is able to identify the forces acting on a given body. It is just a matter of time and more practice before they learn to discard the extra ``fictitious'' force in X.

\subsection{Polarisation : Identifying Forces Among Gender}
One may recall from \textbf{Sec. \ref{Meth}} that our student composition is female-dominated. As such, polarisation may be due to one gender alone; that is, possibly, due to female only. Let us look back at our main graph demonstrating polarisation namely, \textbf{Fig. \ref{fig_pol_percent}}, and see if simple accounting supports this suspicion. At a glance, the wide separation between the solid lines and dashed lines both for pretest and posttest makes it highly unlikely unless otherwise, females far outnumber males which is not the case here; it is only 56\% to 44\%. The difference in the proportion of males and females (56\% - 44\% = 12\%) is incommensurate with the wide gaps\footnote{We formalise this notion of ``gap'' as the degree of polarisation below.} in the pair of graphs for pretest and posttest shown in \textbf{Fig. \ref{fig_pol_percent}}; for instance, for question II, we have a gap of 71\% - 29\% = 42\% for the pretest and 83\% - 17\% = 67\% for the posttest.

\begin{figure}[htb!]
    \hspace{-4.5em}
    \includegraphics[scale=0.95]{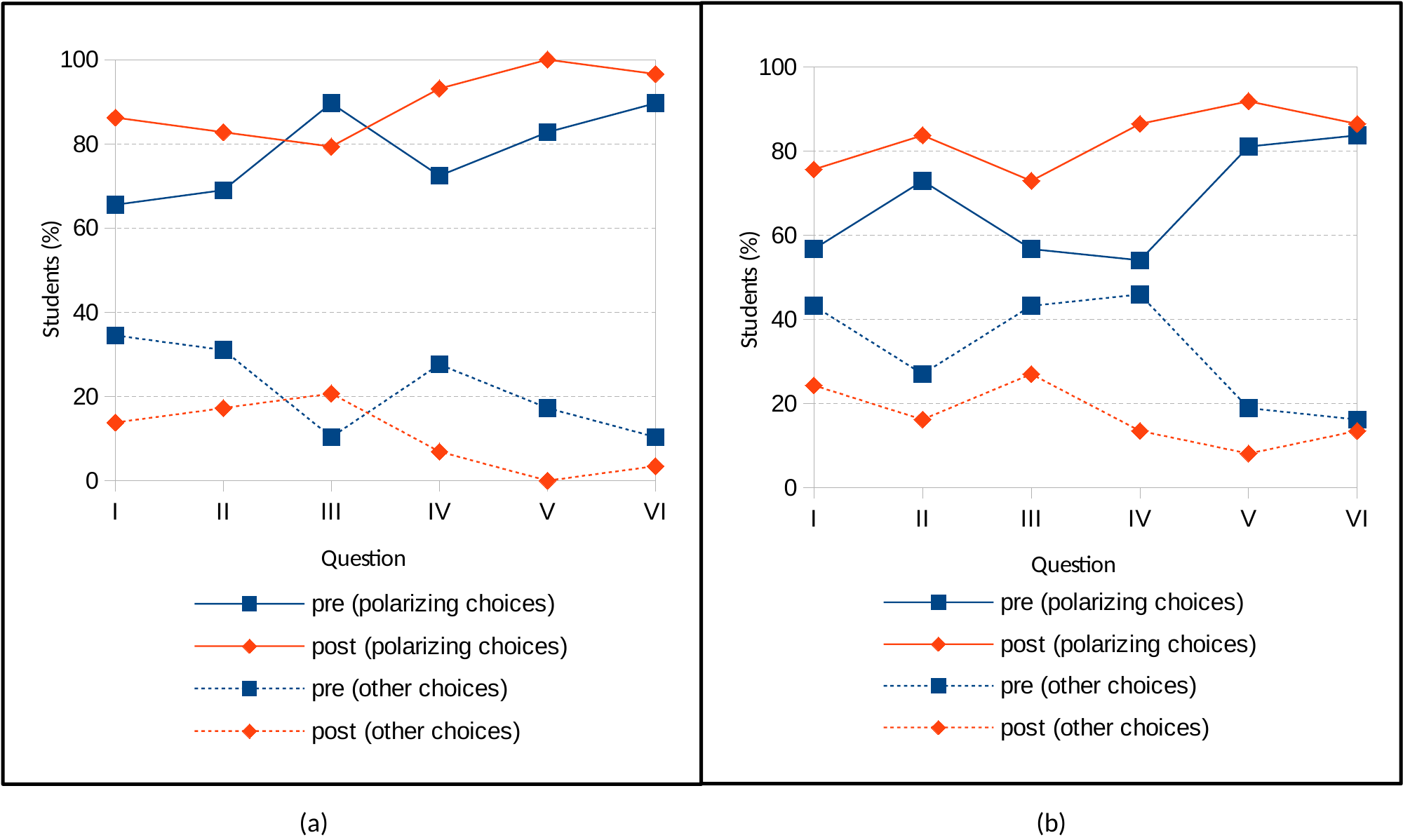}
    \caption{Gender-segregated distribution for polarisation of student answers. The solid lines correspond to the polarising choices while the dashed lines refer to everything else. Moreover, the percent of students on the vertical axis correspond to their respective gender; that is, percent males among males in (a) and percent females among females in (b).}
    \label{genderDifPolarisation}
\end{figure}

The possible gender aspect of polarisation can be seen more clearly by decomposing the graphs in \textbf{Fig. \ref{fig_pol_percent}} into contributions due to males and due to females. \textbf{Figure \ref{genderDifPolarisation}} shows a gender-segregated distribution of student answers. We once again see the same characteristic exhibited in \textbf{Fig. \ref{fig_pol_percent}} indicating polarisation; that is, for each question, there is a \textit{non-vanishing gap} in the pair of graphs for pretest and posttest. Needless to say, this observation holds for both males and females. Coupled with the argument elaborated in the first paragraph above, we then reach one of the main findings of this work--- \textit{polarisation is NOT unique to one gender.} 

Having established this, we now move on to the question about the extent by which it is exhibited by one gender over the other. A visual inspection of the graphs \textbf{(a)} and \textbf{(b)} in \textbf{Fig. \ref{genderDifPolarisation}} seems to indicate that males exhibit a more pronounced polarisation. To shed light on this issue, we introduce a quantitative measure called \textit{degree of polarisation} \textit{P}, defined as the difference between the percent of students who answered the polarising choices and those who elected the remaining choices. It has a minimum value of 0\% indicating no polarisation, and a maximum value of 100\% meaning all students elected the polarising choices. Geometrically, it is a measure of the gap between a pair of line graphs corresponding to polarising choices and other choices, symmetric about the horizontal line at 50\%-mark. (There are two of these pairs in \textbf{Figs. \ref{genderDifPolarisation}(a)} and \textbf{\ref{genderDifPolarisation}(b)} corresponding to pretest and posttest.) The narrower the gaps are, the smaller is the average value of $P$ for all the six questions.

\begin{figure}[htb!]
    \centering
    \includegraphics[scale=0.95]{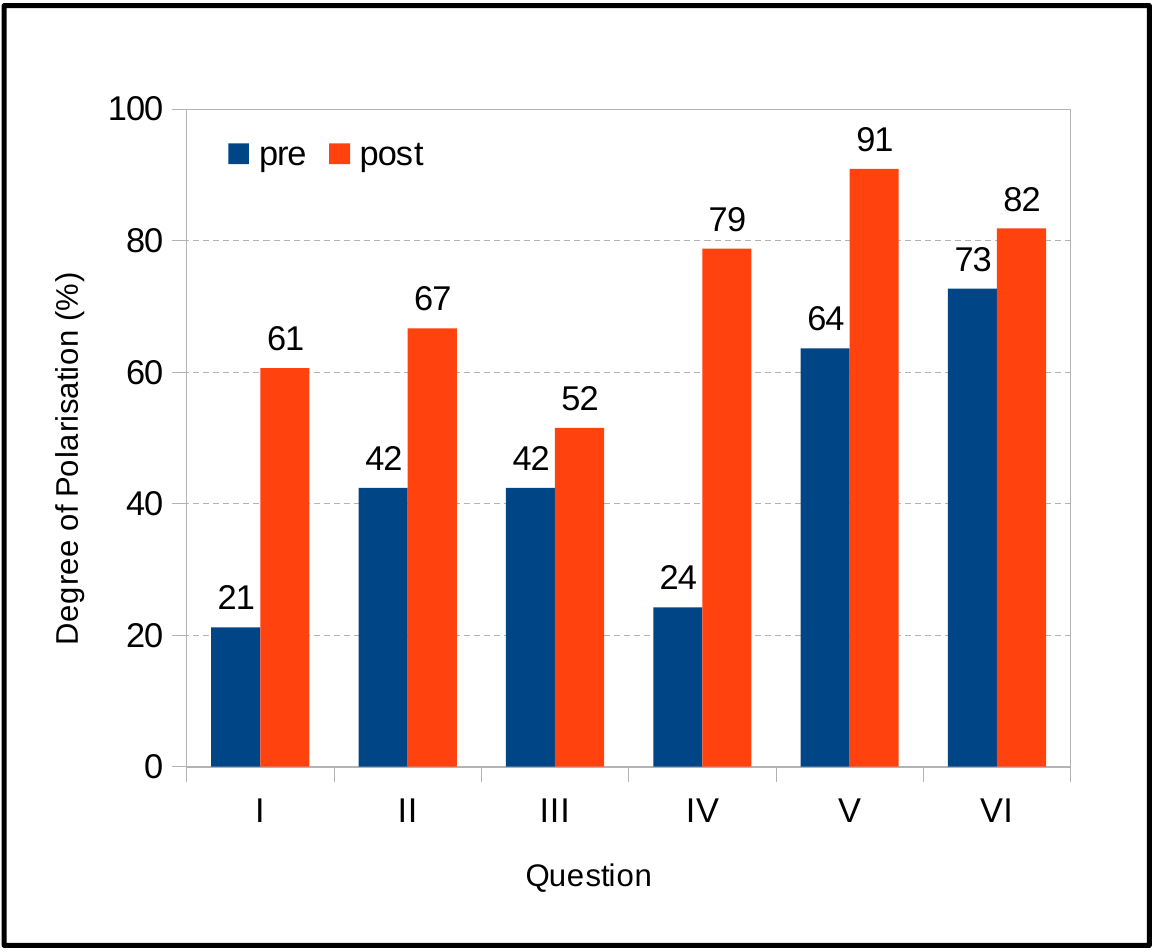}
    \caption{Degree of polarisation for all students on the FCI.}    
    \label{figDegreePolarisation}
\end{figure}

\begin{figure}[htb!]
    \hspace{-4.5em}
    \includegraphics[scale=0.95]{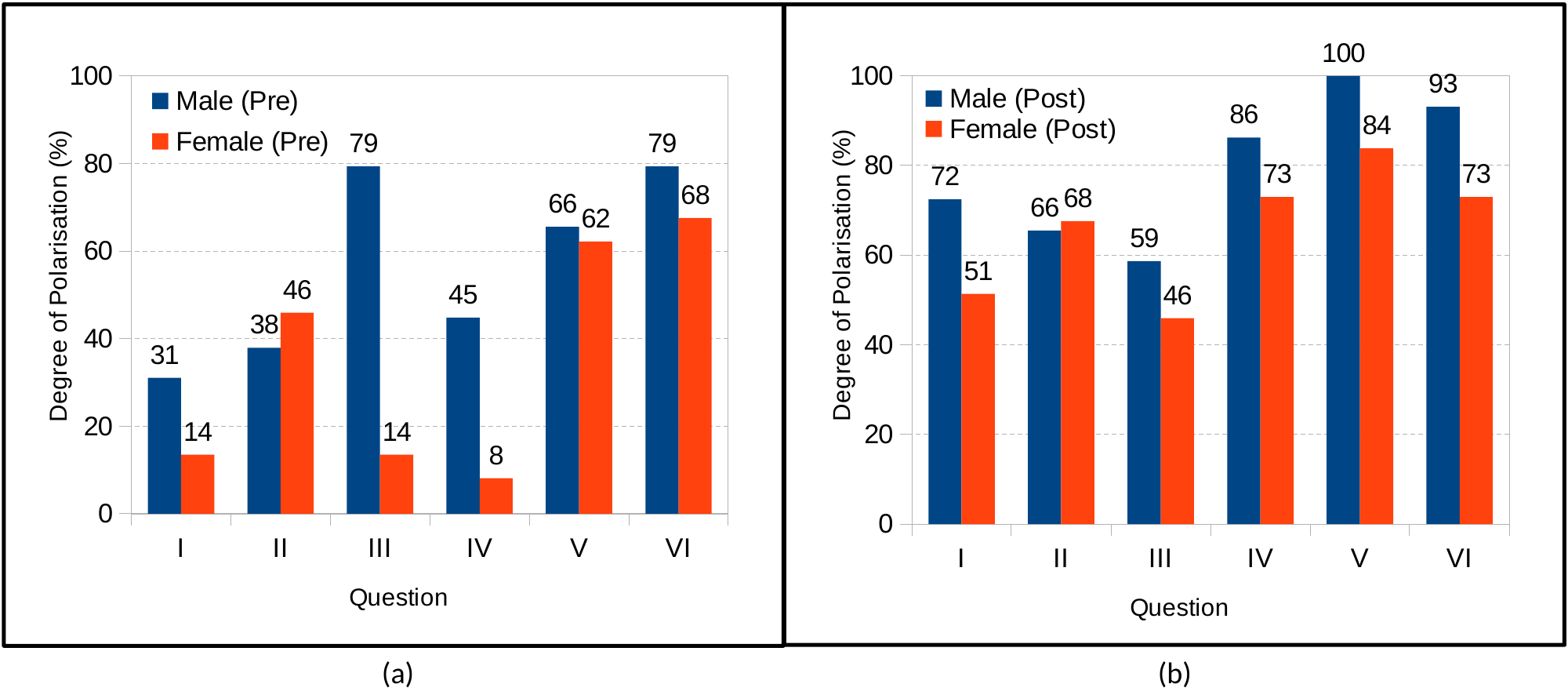}
    \caption{Gender-segregated distribution for the degree of polarisation.}    
    \label{figDegreePolarisationGender}
\end{figure}

\textbf{Figure \ref{figDegreePolarisation}} shows the degree of polarisation for all students irrespective of gender. We see a general pattern that \textit{P} is higher for posttest than for pretest for all the six questions. We split this graph into contributions due to males and due to females (see \textbf{Fig. \ref{figDegreePolarisationGender}}) and then observe the same behaviour with females strictly following the pattern. However, on average, males exhibit a higher degree of polarisation for \textit{both} pretest (56\% versus 35\% for females) and posttest (79\% versus  66\% for females) suggesting a possible gender gap. Such a gap may be interpreted in two ways. (In what follows, keep in mind that on average, males exhibit a higher degree of polarisation.) If a greater proportion of males happen to choose the wrong choice in the pair of polarising choices than that of females, then this may imply that males tend to be more confused. The possible gender gap may then be seen as in favour of females. On the other hand, if a greater proportion of males happen to choose the right choice in the pair of polarising choices than that of females, then this may imply that males tend to be more discerning about the correct set of forces acting on a given body. The possible gender gap may then be seen as in favour of males. 

To resolve this issue, we need to establish if there really is a gender gap associated with the degree of polarisation in the first place. The two (possibly) interesting simulated scenarios we have just presented above can then only take into effect should there be a non-negligible gender difference. This situation seems to demand a quantitative comparison test. A \textit{t}-test involving \textit{P} would be a viable candidate but as we have only six questions corresponding to six pairs of data points for males and females, this would yield a questionable result. On the other hand, one may see that from a slightly different perspective, we are dealing here with the question of independence of two categories namely, (a) gender and (b) choice of answer (polarising choices and other choices). Treating all the polarisation-inducing questions on equal footing, we then find that if we count the frequency of students electing (a) polarising choices and (b) other choices, depending on gender, the appropriate test would be the chi-squared test, and this is what we use here.

For the pretest, our chi-squared test returns a \textit{p}-value of 0.020. This means that at the 95\% confidence level, we have statistically significant evidence that there is a gender aspect associated with polarisation. Looking back at \textbf{Fig. \ref{figDegreePolarisationGender}(a)}, we see that it is not surprising at all. In five out of six questions, the \textit{P}-bars for males are taller than that of females corresponding to an average of 56\% versus 35\%. Further examination of our data indicates that 39\% (40\%) of males (females) got the correct answer in the six questions. It follows that the greater degree of polarisation for males is due to the selection of the choice X in the pair of polarising choices. For the pretest, males seem to be more confused in identifying the correct set of forces acting on a given body. 

For the posttest, our chi-squared test returns a \textit{p}-value of 0.055. This means that at the 95\% confidence level, we do not have statistically significant evidence suggesting a gender aspect of polarisation. With reference to \textbf{Fig. \ref{figDegreePolarisationGender}(b)} we see that although it exhibits the same pattern as that of \textbf{Fig. \ref{figDegreePolarisationGender}(a)}, (that is, in five out of six questions, the \textit{P}-bars for males are taller than that of females,) the relative difference in the degree of polarisation is smaller corresponding to an average \textit{P}-value of 79\% versus 66\%. We may then say that along the way, students (males and females) learned more about forces and the polarisation gender gap observed in the pretest closes at the posttest.

\begin{figure}[htb!]
	\centering
    \includegraphics[scale=0.85]{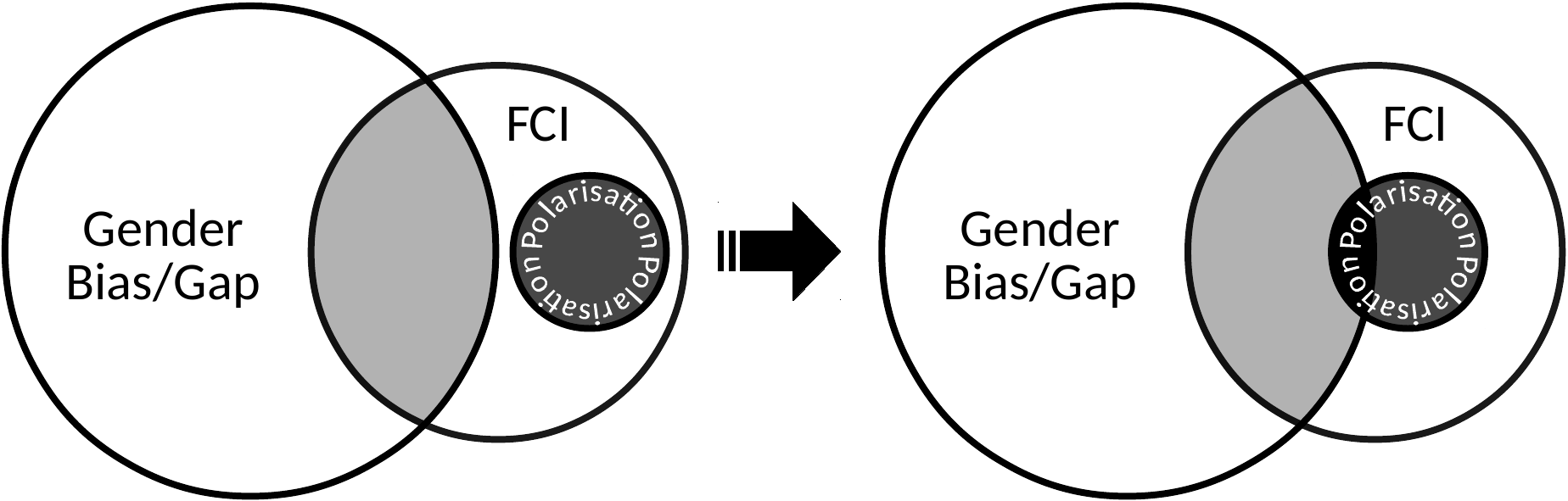}
	\caption{Venn diagram for polarisation, gender bias/gap, and the FCI. Our data suggest that polarisation does not have a serious gender dimension.}    
    \label{figVennExpect}
\end{figure}

We are now in position to go back to the Venn diagram representing the place of polarisation in relation to gender gap and FCI. We have the following main findings:
\begin{itemize}
\item
Polarisation is NOT unique to one gender.
\item
The gender aspect that may be associated with the extent by which polarisation is exhibited by one gender over the other, is not a permanent attribute (if ever it is,) of polarisation.
\end{itemize}
With these two points in hand, we move the circle for polarisation away from that of the gender gap/bias. As far as our results are concerned, it may not be placed completely outside of the zone for gender gap/bias as shown in \textbf{Fig. \ref{figVennExpect}}, indicating that polarisation may have a gender dimension; albeit, something that fades with learning. Whether it will stay in this position (slightly) intersecting the circle for gender bias/gap, remains to be seen in future studies employing larger number of participants.

\section{The FCI as a whole}
\label{ResWhole}
Although this work mainly focuses on the gender aspect of polarisation, it would not be a complete study without mentioning the result of the FCI as a whole. The point in elaborating this is not to repeat what has been found out in the literature; that is, there is gender gap associated with the FCI considered as a whole. In fact, as we shall see, we are part of the minority that deviates from this finding. The additional point then, aside from completeness, is to some extent, a sense of \textit{complementarity}. Our analysis and elaboration below serve to support the idea that polarisation is not a gender-specific phenomenon and the extent by which it is exhibited by students does not have a serious gender aspect.

\subsection{Results and Analyses}

\textbf{Table \ref{table_1}} shows the results of the FCI for four batches of entrants (from AY 2011-12 to AY 2014-15)\footnote{This is an updated set of data we previously presented in Ref. \cite{alalinea}.} corresponding to a total of 66 students, for pretest and posttest. The mean scores for the pretest and posttest are 52\% and 62\% respectively, corresponding to a rise of about 20\%. Our \textit{t}-test returns a \textit{p}-value of $ 2.9\times 10^{-8} $. This indicates that at the 95\% confidence level, the null hypothesis was not satisfied and suggests the results do lead to a positive increase in the post scores.

\begin{table}[tbh!]
\centering
    \begin{tabular}{|l|ccccc|}
    \hline
    \multirow{2}{*}{Group} &\small{2011-2012}  &\small{2012-2013} 
    &\small{2013-2014}  &\small{2014-2015} &\small{Total} \\
       &\small{$n=11$}  &\small{$n=18$} 
    &\small{$n=18$}  &\small{$n=19$} &\small{$N=66$} \\
    \hline
    \hline
       & Mean (SD)    & Mean (SD)    & Mean (SD)    & Mean (SD) & Mean (SD)    \\
    Pretest(\%) & 57 (22) & 55 (25) & 49 (22) & 50 (22) & 52 (23) \\
    Posttest(\%) & 67 (23) & 66 (21) & 55 (23) & 62 (19) & 62 (21) \\
    Gain   & 0.24         & 0.23        & 0.12     & 0.25 &   0.20 \\
    \hline
    \end{tabular}
    \caption{\it FCI results (in percent) for four academic years starting from 2011 and ending 2015. The number $ n $ under each academic year indicates the number of students while the quantity SD stands for standard deviation.}
    \label{table_1}
\end{table}

The \textit{t}-test can only indicate whether there is a statistically significant difference between the pretest and posttest scores. A good measure commonly employed in studies like this to quantify the improvement with respect to the score in the pretest, is provided by the \textit{normalised gain}, \textit{G} \cite{Hake}. It is defined as the difference between the posttest and pretest scores divided by the maximum possible rise in score relative to the pretest result. Symbolically,
\begin{equation*}
	\hspace{5.0em} G = {\langle \% S_f\rangle  - \langle \% S_i \rangle \over 100 - \langle \% S_i \rangle },
\end{equation*}
\noindent
where $\% S_f$ and $\% S_i$ are the final and initial scores in percent respectively.

Also included in \textbf{Table \ref{table_1}} are the \textit{G}-values for each AY for the four batches of students. On average, we find a relatively low value of $G = 0.20$ (see Ref. \cite{coletta} for the interpretation of FCI scores). This is in spite of the use of interactive engagement \cite{Hake, freeman, cahyadi} in class in the form of (a) peer discussion and (b) use of MasteringPhysics${}^\textnormal{\footnotesize TM}$ with the guidance of the instructor/teaching assistant. As argued in our previous work \cite{alalinea}, the corrective measure built into MasteringPhysics${}^\textnormal{\footnotesize TM}$ might have contributed in the low normalised gain. In addition to this, the formal contact hours between students and teacher (one meeting per week each lasting for only 1.5 hours, as stated in \textbf{Sec. \ref{Meth}}) might have been not enough to instill mastery of the core concepts in Newtonian mechanics.
  
\subsection{Gender Differences?}

\begin{table}[tbh!]
\centering
    \begin{tabular}{|l|ccc|}
    \hline
    \multirow{2}{*}{Group} & Female  & Male  & Percent \\
       & $n=37$ & $n=29$ & Difference  \\
    \hline
    \hline
       & Mean (SD)    & Mean (SD)  &  \\
    Pretest(\%) & 49 (22) & 57 (22) & 16  \\
    Posttest(\%) & 58 (21) & 67 (21) & 15\\
    Gain   & 0.18         & 0.23 & 24    \\
    \hline
    \end{tabular}
    \caption{\it FCI results (in percent) for female and male participants from AY 2011-12 to AY 2014-15.}
    \label{table_2}
\end{table}

\textbf{Table \ref{table_2}} is a summary table comparing the performance of males and females in the pretest and posttest. Our data suggest that considering the mean scores, males tend to perform better on \textit{both} the pretest and posttest. This observation is further supported by a higher \textit{G}-value indicating a greater improvement on the part of the males. Looking at the table, a difference of $\Delta G = 0.23 - 0.18 = 0.05$ might not seem much but compared to the average $(\bar G = 0.21)$, this leads to a relatively large percent difference of $24\%$. Having laid down the means and \textit{G}-value, let us see if our \textit{preliminary insight} can survive rigorous statistical test.

\textit{Pretest}. At the $95\,\%$ confidence level, or $\alpha=0.05$, we find that $p = 0.13>\alpha$, so there is no significant difference between the means of the pretest scores of males ($\bar x_{m,pre} = 57\,\%$) and females ($\bar x_{f,pre} = 49\,\%$). Though the percent difference between the means may not be ignored at $16\,\%$ in favour of males, the relatively wide variation of student scores (as measured by the standard deviation) results to acceptance of the null hypothesis in the $t$-test.

\textit{Posttest}. At the $95\,\%$ confidence level, or $\alpha=0.05$, we find that $p = 0.087>\alpha$, so there is no significant difference between the means of the posttest scores of males ($\bar x_{m,post} = 67\,\%$) and females ($\bar x_{f,post} = 58\,\%$). Though the percent difference between the means may not be ignored at $15\,\%$ (only 1\% away from that of the pretest), the relatively wide variation of student scores (as measured by the standard deviation) results to acceptance of the null hypothesis in the $t$-test.

It is worth noting that our \textit{preliminary insight} above is in agreement with the review article by Madsen, McKagan, and Sayre \cite{MadsenRev} on gender gap involving the FCI to the extent that male average score and \textit{G}-value are higher than that of female. However, in our case, the difference in means in favour of males is not supported by the \textit{t}-test results; the fluctuation in student scores is wide enough preventing us to conclude that there is indeed a gender gap. 

Recalling the gender distribution of the participants mentioned in \textbf{Sec. \ref{Meth}}, we find that it is female-dominated (56\% vs. 44\%). In view of Ref. \cite{Shapiro}, the absence of ``stereotype threat'' might have been one of the possible factors leading to a statistically null gender difference.\footnote{This sort of justification seems unnecessary as far as the null result is concerned. However, in view of the overwhelming data on gender gap presented by Madsen, McKagan, and Sayre \cite{MadsenRev}, we find it good to find some words to somehow support our result.}  Furthermore, the fact that majority of our students are international students, their global mindset might correspond to the absence of the gender dimension. Although these are interesting angles to look at, owing to the limited space for this work, we leave the investigation for future studies. 

Our result above brings us back to the gender aspect of polarisation. Had there been a statistically significant gender difference in the FCI performance as a whole, polarisation would have been expected to follow suit. In theory (with polarisation-inducing questions constituting 20\% of all the questions), it is possible that the gender aspect of the two to be opposite each other (one with gender gap while the other has none). What we have found out is that both genders exhibit it. Moreover, the gender aspect associated with the degree of polarisation seems to fade naturally as students learn more about forces. Except for the possible preliminary gender difference, the  null result for the FCI as a whole complements our main findings about polarisation.

\section{Concluding remarks}
\label{Conc}

Polarisation is a phenomenon wherein the addition of a ``fictitious'' force to an otherwise correct set of forces causes a serious divide in the way students identify the valid set of forces acting on a given body. In this article, we have revisited this notion, this time, in relation to gender gap. Our data suggest that polarisation is \textit{not} unique to one gender. Furthermore, the extent by which it is exhibited by males may differ at the beginning from that of females but the gap closes upon learning more about forces. 

These findings involving the six polarisation-inducing questions in the FCI are complemented by our result for the FCI as a whole. We have discussed in this work evidence suggesting that there is no gender gap for global courses in conceptual physics. Although preliminary insight based on student means appeared to show a gender gap, using the \textit{t}-test, we found that the null hypothesis was satisfied at the 95\% confidence level; meaning, there is no statistically significant gender gap.

Our null result for the gender aspect of polarisation and FCI as a whole may require no justification at all. However, in view of the existing evidences to the contrary for the FCI as a whole (see our references), it  might be worth mentioning that our student participants were female-dominated and multi-racial in nature (international students). How these factors might have possibly influenced our result remains to be investigated.

As for future work, it is apparent that we would need a larger data set. Moreover, it would be beneficial to look at other international/global courses that have other interesting cohorts (not just the CBCMP program at Osaka University). Certainly, the study of whether or not `stereotype threats' \cite{Shapiro}  affect performance in the international context warrants further investigation. 

\section*{Acknowledgements}

We would like to thank current and past students and staff on the
Chemistry-Biology Combined Major Program (CBCMP) at Osaka 
University.


\end{document}